\documentclass[pra,dvips,showpacs]{revtex4}
\usepackage[dvips]{graphicx}

\usepackage{amsfonts}
\usepackage{amssymb}
\usepackage{bbm}

\newcommand{\one}{\mbox{$1 \hspace{-1.0mm}  {\bf l}$ }}

\begin{document}

\title{Berry phase in open quantum systems: a quantum Langevin equation approach}

\author{G. De Chiara}
\affiliation{ NEST- INFM \& Scuola Normale Superiore, piazza dei Cavalieri 7 , I-56126 Pisa, Italy}
\author{ A. {\L}ozi\'nski }
\affiliation{ NEST- INFM \& Dipartimento di Tecnologie dell'Informazione, Universit\`{a} degli Studi di Milano; via Bramante 65,
I-26013 Crema (CR), Italy}
\author{G. M. Palma}
\affiliation{  NEST- INFM \& Dipartimento di Scienze Fisiche ed Astronomiche, Universit\`{a} degli Studi di Palermo, via Archirafi
36, I-90123, Palermo, Italy }

\begin{abstract} The evolution of a two level system  with a slowly varying Hamiltonian, modeled as a spin 1/2 in a slowly varying
magnetic field, and interacting with a quantum environment, modeled as a bath of harmonic oscillators is analyzed using a quantum
Langevin approach. This allows to easily obtain the dissipation time and the correction to the Berry phase in the case of an
adiabatic cyclic evolution.
\end{abstract}


\maketitle

%\date{}

\section{Introduction}
In his seminal work \cite{Berry} Berry showed the appearance of a purely geometrical phase factor associated to the non degenerate
eigenstates of a Hamiltonian undergoing a cyclic adiabatic evolution. Since then much work has been done to generalize such
concept to non cyclic, non degenerate or non adiabatic evolution \cite{Shapere}. The renewed interest for geometric phases in a
quantum computation scenario \cite{Jones,Zanardi,Falci,Zoller,Ekert} is due to their supposed intrinsic fault tolerance. Such
hypothesis has been analyzed in \cite{DeChiara} for the case of a spin in a slowly varying magnetic field with small classical
random fluctuations. There it was shown that for small fluctuations, i.e. to first order in the perturbation, and in the adiabatic
limit the main source of decoherence are dynamical fluctuations. Similar conclusions have been reached for quantum noise in
\cite{Carollo} by means of a quantum trajectories approach. In the present paper we will analyze the problem using a quantum
Langevin equation approach. Our system consists of a pseudospin interacting with a quantized bosonic field. The spin free
Hamiltonian is assumed to undergo a slow cyclic evolution. The geometric phase appears in a natural way in terms of the so called
adiabatic Hamiltonian \cite{adiabatic}. Once such Hamiltonian is introduced the Heisenberg equations of motion are derived and
from them the quantum Langevin equations. This approach allowed us to analyze the effects of the quantum fluctuations on both the
decay constants and on the overall phase acquired by the spin energy eigenstates in their cyclic evolution. The same problem has
been addressed by some recent papers \cite{Witney} with the use of an elaborated perturbative technique. We will show how our
approach allows to obtain in a simpler and more straightforward way the corrections to the the Berry phase found in \cite{Witney}.
Furthermore we will provide a transparent physical picture of the results obtained. The effect of noise on geometric phases in
different scenarios from the one described above has been studied in \cite{Blais}.

\section{The adiabatic Hamiltonian}
\label{se_ad_app}

The system we consider consists of a pseudospin in a slowly varying static magnetic field interacting with  an
environment modeled as a bath of harmonic oscillators. The overall system Hamiltonian is assumed to be of the
standard form
\begin{equation}
\hat{H} = \frac{1}{2} {\bf B}\cdot{\vec \sigma} + \sum_k \omega_k \hat{a}_k^{\dagger} \hat{a}_k +\sum_k g_k \sigma_z\left(
\hat{a}_k + \hat{a}^{\dagger}_k \right)
\end{equation}
where $\vec{\sigma} \equiv (\sigma_x,\sigma_y,\sigma_z)$ are Pauli operators and ${\bf B}(t) \equiv B_0 ( \sin\vartheta
\cos\varphi , \sin\vartheta \sin\varphi , \cos\vartheta )$ is a three dimensional vector, which we assume to be time dependent and
$\hat{a}_k^{\dagger} ( \hat{a}_k)$ are bosonic creation (annihilation) operators for mode $k$ (we have set $\hbar =1$).

The first step to obtain the Heisenberg equation of motion for the spin and bath operators is the introduction of the so called
adiabatic Hamiltonian \cite{adiabatic} i.e. of the Hamiltonian whose eigenstates, in the absence of interaction with the
environment, after a cyclic evolution acquire the dynamical and geometrical phase predicted by Berry. Let us first rewrite the
free spin Hamiltonian in the form

\begin{equation}
\hat{H}_S \equiv \frac{1}{2} {\bf B}\cdot{\vec \sigma} = \frac{B_0}{2} \left( |\uparrow_n\!(t)\rangle\langle \uparrow_n\! (t)| -
|\downarrow_n\! (t)\rangle\langle\downarrow_n\! (t)| \right)
\end{equation}
where $|\uparrow_n\!(t)\rangle$ and $|\downarrow_n\!(t)\rangle$ are the eigenstates of $\hat{H}_S$ at time $t$ i.e.  the
eigenstates of the operator $ {\vec \sigma}\cdot{\bf n}$, where $\bf n\equiv (\sin\vartheta \cos\varphi , \sin\vartheta
\sin\varphi , \cos\vartheta )$ is a unit vector pointing in the instantaneous ${\bf B}$ direction. Let us then define the
following time dependent unitary operator:

\begin{equation}
U(t) = |\uparrow_n\! (0)\rangle\langle\uparrow_n\!(t)| + |\downarrow_n\!(0)\rangle\langle\downarrow_n\!(t)|
\end{equation}
In the absence of any coupling with the environment the time evolution of the state vector $|\tilde{\psi}(t)\rangle = \hat{U}(t)
|\psi(t)\rangle$ is generated by the Hamiltonian

\begin{equation}
\hat{\tilde{H}}_S = \hat{U}(t)\hat{H}_S \hat{U}^{\dagger}(t) - i\hat{U}(t)\frac{d}{d t} \hat{U}^{\dagger}(t) \label{h2}
\end{equation}
A time dependent $\bf B$ will in general induce transitions between the instantaneous energy eigenstates $|\uparrow_n\!
(t)\rangle$ and $ |\downarrow_n\!(t)\rangle$. However if the direction of $\bf B$ changes slowly enough in time we can neglect
such transitions, which in (\ref{h2}) are described by the term $i\langle\uparrow\!(t)|\frac{d}{d t}|\downarrow(t)\rangle $. This
is nothing but the standard adiabatic approximation \cite{Messiah}, valid as long as $i\langle\uparrow\!(t)|\frac{d}{d
t}|\downarrow(t)\rangle \ll B$. On the other hand $i\langle\uparrow\!(t)|\frac{d}{d t}|\uparrow(t)\rangle $ and
$i\langle\downarrow\!(t)|\frac{d}{d t}|\downarrow(t)\rangle $, which are nothing but the so called Berry connection, are
responsible of the appearance of a non vanishing geometric phase and therefore must not be neglected. The adiabatic Hamiltonian is
therefore

\begin{eqnarray}
\hat{H}^{ad}_S =\left( \frac{B_0}{2}-i\langle \uparrow_n\!(t)| \frac{d}{d t} |\uparrow_n\!(t)\rangle\right)
|\uparrow_n\!(0)\rangle\langle \uparrow_n\! (0)| - \left( \frac{B_0}{2}+i\langle \downarrow_n\!(t)| \frac{d}{dt} |\downarrow_n
\!(t)\rangle\right) |\downarrow_n\! (0)\rangle\langle\downarrow_n\! (0)|  \label{Had}
\end{eqnarray}
Note that when $\bf{B}$ undergoes a cyclic evolution the eigenstates of (\ref{Had}) correctly acquire the dynamical plus the
geometrical phase predicted by Berry.

In the basis of the eigenstates of the $\sigma_z$ operator we have

\begin{eqnarray}
|\uparrow_n \rangle &=& e^{-i\varphi/2} \cos \frac{\vartheta }{2}|\uparrow_z \rangle
+ e^{i\varphi/2}\sin \frac{\vartheta }{2}|\downarrow_z \rangle\nonumber\\
|\downarrow_n \rangle &=& e^{-i\varphi/2}\sin \frac{\vartheta}{2}|\uparrow_z \rangle- e^{i\varphi/2}\cos \frac{\vartheta }{2}|
\downarrow_z\rangle
\end{eqnarray}
from which it follows that

\begin{equation}
i\langle \uparrow_n\!(t)| \frac{\partial}{\partial t} |\uparrow_n\!(t)\rangle = - i\langle \downarrow_n\!(t)|
\frac{\partial}{\partial t} |\downarrow_n \!(t)\rangle = \dot{\varphi}\frac{1}{2}\cos\vartheta
\end{equation}
The adiabatic Hamiltonian takes therefore the form

\begin{equation}
\hat{H}^{ad}_S = \frac{B_0 - \dot{\varphi}\cos \vartheta (t) }{2}{\bf n}(0)\cdot {\vec \sigma}
\end{equation}

For the sake of simplicity, and with no loss of generality, we will consider the case most discussed in literature in which ${\bf
B} (t)$ precesses slowly  with angular velocity $\Omega = 2\pi/T$ i.e. $\varphi (t) = \Omega t , \varphi (0) = 0, \vartheta (t) =
\vartheta (0)$. Furthermore we will rotate our axis so that $[ \cos\vartheta (0) \sigma_z + \sin\vartheta (0)\sigma_x ]
\rightarrow\sigma_z$ In this case the adiabatic Hamiltonian takes the following simple form

\begin{equation}
\hat{H}^{ad}_S = \frac{\omega_0}{2} \sigma_z
\end{equation}
where $\omega_0 = B_0 - \Omega\cos \vartheta (0)$. In order to simplify the comparison between our results and the existing
literature we would like to point out that our approach is different from, although at the end equivalent to, the one in which a
rotating frame is introduced (see.e.g. \cite{Witney}). In geometric terms this implies that a field of amplitude $\Omega $ along
the $\hat{z}$ direction is added to ${\bf B}$, i.e. in the rotating frame the effective magnetic field changes both in direction
and length. This is not the case in our adiabatic Hamiltonian, in which the magnetic field changes only in length. However when
the overall accumulated phase is calculated to lowest order in $\Omega$ their model coincides with ours. In other words while we
assume from the very beginnings the adiabatic limit in \cite{Witney} the non adiabatic contributions are discarded a posteriori.

To conclude this section we write our full, time independent,
adiabatic spin - boson Hamiltonian
\begin{eqnarray}\label{hamiltonian}
  {\hat H}^{ad} =\frac{\omega_{0}}{2}\sigma_z + \sum_k \omega{}(k) \hat{a}^{\dagger}_k\hat{a}_k + \sum_k g_k \left( \hat{a}_k +
\hat{a}^{\dagger}_k \right)\left( \sigma_z\cos\vartheta - \sigma_x\sin\vartheta \right)
\end{eqnarray}

Note that in (\ref{hamiltonian}) and from now on the spin operators are in the adiabatic frame.

\section{The Langevin equations}
\label{se_la_eq}

The quantum Langevin equation of motion for a two level systems coupled with a harmonic bath is well known in literature
\cite{CCT,Eberly}. Here, for the sake of clarity, we will sketch the main steps of its derivation. The starting point are the
Heisenberg equations of motion ($\dot{O} = i[\hat{H},O]$) for the spin Pauli operators $\sigma_z$, $\sigma_{+} = (\sigma_x +
i\sigma_y)/2$ and the bath operators ${\hat{a}}_k , {\hat{a}}^{\dagger}_k$ which are
\begin{eqnarray}
\dot{\sigma}_z &=& 2i\sin\vartheta\sum_k g_k \left(\sigma_{+} - \sigma_{-}
\right) \left( \hat{a}_k + \hat{a}^{\dagger}_k \right)  \label{Heisenberg1}\\
\dot{\sigma}_{+} \!&=&\! i\omega_0\sigma_{+} + i\!\sum_k g_k\!\!\left( \hat{a}_k + \hat{a}^{\dagger}_k \right)\! \left( \sin\vartheta\sigma_z\!
+\! 2\cos\vartheta\sigma_{+} \right) \label{Heisenberg2}
\\
\dot{\hat{a}}_k &=& -i\omega_k\hat{a}_k -ig_k \left( \cos\vartheta\sigma_z - \sin\vartheta(\sigma_{+} + \sigma_{-}) \right)
\end{eqnarray}
which can be cast in the following integral form
\begin{eqnarray}
 \sigma_z(t)   &=&\sigma_z(0) + 2i \sin\vartheta \sum_k g_k \int_0^t dt'\left[ \sigma_{+}(t') - \sigma_{-}(t') \right] \left[
\hat{a}_k(t')
+\hat{a}^{\dagger}_k(t') \right]   \label{eq_for_sol_1} \\
 \sigma_{+}(t) &=&\sigma_{+}(0)e^{i\omega_0t} + i\sum_k g_k \int_0^t dt' e^{i\omega_0(t-t^{'})} \left[ \hat{a}_k(t') +
\hat{a}^{\dagger}_k(t') \right]
\left[ \sin\vartheta\sigma_z(t') + 2\cos\vartheta\sigma_{+}(t') \right]
 \label{eq_for_sol_2} \\
 \hat{a}_k(t)  & =&\hat{a}_k(0)e^{-i\omega_k t} -ig_k\int_0^t dt' e^{-i\omega_k (t-t^{'})} \left\{ \cos\vartheta \sigma_z(t')\sin\vartheta\left[ \sigma_{+}(t')+\sigma_{-}(t')\right] \right\} \label{eq_for_sol_3}
\end{eqnarray}
A standard assumption in the derivation of a quantum Langevin equation is that the timescale of the decay processes is much slower
than the free evolution. In other words the weak coupling with the bath degrees of freedom modifies the spin dynamics on
timescales which are much longer that $\omega_0^{-1}$. In the above integrals we can therefore put
\begin{eqnarray}
\sigma_z(t^{'}) &=& \sigma_z(t) \label{eq_ad_app1}\\
\sigma_{+}(t^{'}) &=& e^{-i\omega_0(t-t^{'})}\sigma_{+}(t) \label{eq_ad_app2}\\
\hat{a}_k(t^{'}) &=& e^{i\omega_k (t-t^{'})}\hat{a}_k(t) \label{eq_ad_app3}
\end{eqnarray}

Eqs.~(\ref{eq_for_sol_1}, \ref{eq_for_sol_2}, \ref{eq_for_sol_3})
then become
\begin{eqnarray}
  \sigma_z &=& \sigma_z(0) + \sin\vartheta \sum_k g_k \left[ \sigma_{+}\hat{a}_k\zeta^{\star} (\omega_0-\omega_k) +\sigma_{+}\hat{a}^{\dagger}_k\zeta^{\star}(\omega_0+\omega_k)
   - \sigma_{-}\hat{a}_k\zeta(\omega_0+\omega_k)- \sigma_{-}\hat{a}^{\dagger}_k\zeta (\omega_0-\omega_k)
\right]  \label{slow 1}\\
 \sigma_{+} &=& e^{i\omega_0t}\sigma_{+}(0) + i\sin\vartheta \sum_k g_k \sigma_z \left[ \hat{a}_k\zeta(\omega_0+\omega_k) +
\hat{a}^{\dagger}_k\zeta(\omega_0-\omega_k) \right] +  2i\cos\vartheta \sum_k g_k \sigma_{+} \left[ \hat{a}_k\zeta(\omega_k) +
\hat{a}^{\dagger}_k\zeta^{\star} (\omega_k)
\right]  \label{slow 2}\\
 \hat{a}_k &=& e^{-i\omega_k t}\hat{a}_k(0)- ig_k\left[ \cos\vartheta \sigma_z \zeta^* (\omega_k) - \sin\vartheta \left(
\sigma_{+}\zeta^{\star}(\omega_0+\omega_k) + \sigma_{-}\zeta(\omega_0-\omega_k) \right) \right] \label{slow 3}
\end{eqnarray}
where
\begin{equation}
\zeta(x) = \lim_{t\to \infty}\int_0^te^{ixt^{'}}dt^{'} = P\frac{i}{x} + \pi \delta(x) \label{zeta}
\end{equation}
$P$ denotes principal part and $\delta (x)$ is the Dirac delta
function. The integration limit $t\rightarrow \infty$ is justified
on the ground that we are interested at times $t\gg
\omega_0^{-1}$.

Inserting the (\ref{slow 1} - \ref{slow 3}) into the (\ref{Heisenberg1} - \ref{Heisenberg2}) and taking care of a consistent choice
of all operator products \cite{CCT,Eberly}, one obtains the desired equations of motion for the spin operator in the rotating frame,
averaged over the environment degrees of freedom

\begin{eqnarray}
  \frac{d\sigma_z }{dt} &=& -2\sin^2\vartheta
 \left(\gamma_{\perp}\sigma_z + \gamma_{\perp vac}\one
\right) - 2\sin 2\vartheta\gamma_{\parallel}\left(
\sigma_{+}+\sigma_{-}\right)  \\
 \frac{d\sigma_+ }{dt} &=& i\omega_0\sigma_{+} + \frac{\sin 2\vartheta}{2}[2(i\xi -\gamma_{\perp vac})\one
-(i\lambda + \gamma_{\perp}) \sigma_z  ] + \sin^2\vartheta \{(i\lambda - \gamma_{\perp}) \sigma_{+} - (+i\lambda  +
\gamma_{\perp})\sigma_{-}\}- 4\cos^2\vartheta \gamma_{\parallel}\sigma_{+}
\end{eqnarray}

 where, transforming the sums into integrals
\begin{eqnarray}
  \gamma_{\perp}  & = & \pi\int^{\omega_c}_0d\omega_k \rho (\omega_k)g_k^2 \left(2n_k+1\right)
\delta(\omega_0-\omega_k) \\
 \gamma_{\parallel} & = &
\pi\int^{\omega_c}_0d\omega_k \rho (\omega_k)g_k^2 \left(2n_k+1\right)\delta(\omega_k)\\
 \lambda\! &=&\!\! \int^{\omega_c}_0\!\!d\omega_k \rho (\omega_k)g_k^2 (2n_k+1)\!\left(\!\frac{P}{\omega_0 -
\omega_k}+ \frac{P}{\omega_0 + \omega_k}\!\right)\\
 \xi\! &=&\!\! \int^{\omega_c}_0\!\!d\omega_k \rho (\omega_k)g_k^2\! \left[ \left( \frac{2P}{\omega_k} - \frac{P}{\omega_0 - \omega_k} +
\frac{P}{\omega_0 + \omega_k} \right) \right].
\end{eqnarray}
In the above equations $ \rho (\omega_k)$ is the density of modes at frequency $\omega_k$, $n_k$ is the mean number of photon in
field mode $k$, $\one $ is the identity in $\mathbb{C}^2$ and $\gamma_{\perp vac}$ is $\gamma_{\perp}$ for $n_k = 0$ .

\section{Dissipation and energy shifts}
\label{se_geph}

The above equations allow us to clearly identify the effects of
the adiabatic evolution on the physical quantities which
characterize the spin dynamics, namely the decay constants and the
energy shift. First of all let us consider the decay constant
$\gamma_{\parallel}$ which describes the decoherence mechanism due
to fluctuations "parallel" to the instantaneous direction of $\bf
B$. As expected it is not modified by the adiabatic change of such
direction. Furthermore the value of $\gamma_{\parallel}$  depends
on the density of field modes at zero frequency which, in most
situations of physical interest is equal to zero.

The decay constant $\gamma_{\perp}$ describes the dissipation mechanism due to the exchange of energy between system and bath and
depends on the density of modes at the resonance frequency $ \omega = \omega_0$. If we assume that the density of modes is a slowly
varying function of $\omega$ near resonance, we can safely assume for very small $\Omega$, i.e. in the adiabatic limit,
$\int^{\omega_c}_0d\omega_k \rho (\omega_k)g_k^2 \left(2n_k+1\right) \delta(\omega-\omega_0) \approx \int^{\omega_c}_0d\omega_k \rho
(\omega_k)g_k^2 \left(2n_k+1\right) \delta(\omega - B_0)$. This confirms that the timescale of dipole decay is not modified by the
adiabatic evolution, a result which has been obtained with different techniques, from classical stochastic noise \cite{DeChiara}, to
quantum jump \cite{Carollo}. We should point out that in order to observe the geometric phase we must have

\begin{equation}\label{eq_timescale2}
\gamma_{\perp}\ll \Omega \ll \omega_0
\end{equation}

Let us consider now the change in the energy shift $\lambda$. In
the adiabatic limit we must consider terms up to order $O(\Omega
)$ and therefore

\begin{eqnarray}
\lambda  & = &\int^{\omega_c}_0d\omega_k \rho (\omega_k)g_k^2 (2n_k+1)\left[\left( \frac{P}{B_0 - \omega} + \frac{P}{B_0 +
\omega}\right) - \Omega \cos\vartheta\left.\frac{\partial}{\partial \omega_0}\right|_{\omega_0 = B_0} \left( \frac{P}{\omega_0 -
\omega}+ \frac{P}{\omega_0 + \omega}\right) \right]\\ &\approx &\lambda_0 + \delta \lambda
\end{eqnarray}
The quantity $\sin^2 \vartheta\lambda_0$ is nothing but the Lamb Shift \cite{CCT,Eberly}, while

\begin{eqnarray}
 \delta \lambda &=&  \Omega \cos\vartheta\int^{\omega_c}_0d\omega_k \rho (\omega_k)g_k^2 (2n_k+1)\left[ \frac{1}{(B_0 - \omega
)^2} + \frac{1}{(B_0 + \omega )^2}\right]
\end{eqnarray}
gives information on the the effect of the quantum fluctuations on the geometric phase. This correction coincides with the results
obtained by \cite{Witney} with an elaborated perturbation technique. The observable overall phase difference between the two energy
eigenstates at the end of their cyclic evolution, i.e. at time $T= 2\pi\Omega^{-1}$, will be

\begin{equation}
\Phi (T) = \Phi_{D} + \Phi_{G}
\end{equation}
where the dynamical phase  $\Phi_{D}$
\begin{eqnarray}
 \Phi_{D}  &=& \left[B_0 + \sin^2\vartheta \int^{\omega_c}_0d\omega_k \rho (\omega_k)g_k^2 (2n_k+1) \left( \frac{P}{B_0 - \omega}+
\frac{P}{B_0 + \omega}\right)\right]T \label{phiD}
\end{eqnarray}
is simply due to the renormalized energy splitting, while the
geometric phase $\Phi_{G}$ is
\begin{eqnarray}
 \Phi_{G}  &=& 2\pi \cos\vartheta \left\{ 1 - \sin^2\vartheta\int^{\omega_c}_0d\omega_k \rho (\omega_k) g_k^2 (2n_k+1)
\left[ \frac{1}{(B_0 - \omega )^2} + \frac{1}{(B_0 + \omega )^2}\right]\right\} \label{phiG}
\end{eqnarray}
The expression (\ref{phiG}) is amenable to a straightforward intuitive geometric interpretation. It is a well known fact that for a
spin $1/2$ the Berry phase is is equal to the solid angle spanned by the time varying magnetic field $\bf B$ on a unit sphere
centered around degeneracy. As opposite energy eigenstates acquire opposite geometric phases the overall phase difference between
them will be, for a slowly precessing field at an angle $\vartheta$, equal to $\Phi_{Berry} = 2\pi\cos\vartheta$. In the presence
of a weak coupling with the bosonic bath however each energy eigenstate will undergo  virtual transitions, responsible for the
Lamb Shift, with a probability
\begin{eqnarray}
  Prob_{vt} &=& \sin^2\vartheta\int^{\omega_c}_0d\omega_k \rho (\omega_k)g_k^2 (2n_k+1)\left[ \frac{1}{(B_0 - \omega )^2} +
\frac{1}{(B_0 + \omega )^2}\right]
\end{eqnarray}
During such transition the spin state parallel (antiparallel) to the direction of the field $\bf B$ 'jumps' to the antiparallel
(parallel) spin state, acquiring an opposite geometric phase. The overall geometric phase difference between the energy
eigenstates will be therefore decreased by an amount proportional to $Prob_{vt}$, as shown in (\ref{phiG}). Notice that the
correction to the Berry phase is of order $O(g^2)$. In \cite{DeChiara}, where the effects of classical noise were considered,  no
analogous correction was obtained because only contributions to first order in the fluctuating field were taken into account.

\section{Conclusions}
In this paper we have shown how the corrections to the Berry phase
and the decay constants for a spin $1/2$ undergoing an adiabatic
cyclic evolution can be obtained in terms of quantum Langevin
equation once the adiabatic Hamiltonian is introduced. We have
confirmed that the main source of decoherence is due only to the
dynamical fluctuations, a result which has been obtained with
different techniques, from classical stochastic noise
\cite{DeChiara}, to quantum jump \cite{Carollo} and which emerges
in a straightforward way in our approach. The Heisenberg equations
of motion give also the correction to the geometric part of the
overall phase difference between the energy eigenstates at the end
of the cyclic evolution due to the coupling with the bath. We have
also shown how such corrections are amenable to a straightforward
geometrical interpretation.

%%%%%%%%%%%%%%%%%%%%%%%%%%%%%%%%%%%%%%%%%%%%%%%%%%%%%%%%%%%%%%%%%

\section*{Acknowledgement}
 We would like to thank Prof.~G.~Falci, Prof.~R.~Fazio and Dr.~E.~Paladino, for helpful discussions. This work was
supported in part by the EU under grant IST - TOPQIP, "Topological Quantum Information Processing" Project, (Contract
IST-2001-39215).


\begin{thebibliography}{99}

\bibitem{Berry}
Berry M.V., Proc.Royal Soc. (London) \textbf{392} (1984) 47, reprinted
        in~\cite{Shapere}.

\bibitem{Shapere}
  For an overview of the broad litterature on the field see:
\textit{Geometric phases in physics}
 Shapere A.  and Wilczek F. Eds.
(World Scientific, Singapore 1989)


\bibitem{Jones}
  Jones J.,  Vedral V., Ekert A.K. and C.Castagnoli, Nature \textbf{403} (2000) 869.

\bibitem{Zanardi}
  Zanardi P. and Rasetti M., Phys.Rev. A \textbf{264} (1999) 94.

\bibitem{Falci}
 Falci G., Fazio R., Palma G.M., Siewert J. and
       Vedral V.
       Nature \textbf{403} (2000) 869,  Faoro L., Siewert J., and Fazio R., Phys.Rev.Lett. \textbf{90} (2003) 028301.


\bibitem{Zoller}
   Duan M.-M , Cirac I. and Zoller P., Science \textbf{292} (2001) 1695.

\bibitem{Ekert}
 Ekert A., Ericsson M., Hayden P., Inamori H., Jones J. A., Oi D. K. L. and Vedral V. J., Mod. Opt. \textbf{47} (2000) 2501.


\bibitem{DeChiara}
   DeChiara G. and Palma G.M., Phys.Rev. Lett. \textbf{91} (2003) 090404.


\bibitem{Carollo}
 Carollo  A., Fuentes-Guridi I., Franca Santos M. and Vedral V., Phys.Rev.Lett. \textbf{90} (2003) 090404,
 Phys.Rev.Lett. \textbf{92} (2004) 020402.

\bibitem{adiabatic}
Vidal J. and Wudka J., Phys. Rev. A \textbf{44} (1991) 5383.
  Brihaye Y. and Kosi\'{n}ski P., Phys.Lett. A \textbf{195} (1994) 296.



\bibitem{Messiah}
Messiah A.
  \textit{Quantum Mechanicst}
 (John Wiley, New York 1961)


\bibitem{Witney}
 Whitney R. S.  and Gefen Y., Phys.Rev.Lett. \textbf{90} (2003) 190402.
Whitney R., Makhlin Y., Shnirman A. and Gefen Y., arXiv:cond-mat/0401376,
        arXiv:cond-mat/0405267. (2004


\bibitem{Blais}
Blais A. and Tremblay A.-M.S., Phys.Rev.A \textbf{67}, (2003),012308. Nazir A., Spiller T. P. and Munro W. J., Phys. Rev. A
\textbf{65} (2002) 042303.


\bibitem{keiji}
Wang X.-B. and Matsumoto K., J.Phys.A \textbf{34}(2001), L631, Phys.Rev.Lett.\textbf{87} (2001) 097901

\bibitem{CCT}
 Cohen-Tannoudji C., Dupont-Roc J. and Grynberg G. \textit{Atom-photon interactions : basic processes and applications}
 (John Wiley, New York 1992) Complement $A_V$



\bibitem{Eberly}
Ackeralt  J.R. and Eberly J.H.,
  Phys.Rev.D \textbf{10}( 1974) 3350.


\end{thebibliography}
\end{document}